\documentclass{iopconfser}
\usepackage{amsmath}
\usepackage{amssymb}
\usepackage{graphicx}
\usepackage{hyperref}

\newcommand{\dd}{{\rm{d}}}
\newcommand{\im}{{\rm{i}}}

\begin{document}

\title{A novel class of rotating black holes with non-aligned electromagnetic field}

\author{Hryhorii Ovcharenko$^{1}$, Ji\v{r}\'{i} Podolsk\'{y}$^{1}$}

\affil{$^1$Institute of Theoretical Physics, Charles University, Prague, Czechia}

\email{hryhorii.ovcharenko@matfyz.cuni.cz}

\begin{abstract}
We present a new class of expanding and twisting solutions to the Einstein-Maxwell equations of algebraic type D, where the null eigendirections of the Faraday tensor are not aligned with PNDs of the Weyl tensor. After deriving this novel solution, we explore its various metric forms and parameterizations. In suitable coordinates, the solution depends on six physical parameters, namely mass $m$, Kerr and NUT twist parameters $a$ and $l$, complex charge $c$, acceleration $\alpha$, and parameter $\beta$ that governs the interplay between electric and magnetic charges in the aligned part of the Faraday tensor. This parameterization, as the Griffiths-Podolsk\'{y} form of the Pleba\'{n}ski-Demia\'{n}ski solution, facilitates explicit special subcases, such as Kerr-Newman black holes, and a deeper physical interpretation. Additionally, in the static limit, our solution reduces to previously known cases.
\end{abstract}

\section{Introduction}
Black holes are extraordinary theoretical predictions originating from Einstein's general relativity. Recent support for their existence has emerged through the detection of gravitational waves resulting from the merging of binary black holes, as well as the images capturing the shadows of supermassive black holes in M87* and in Sgr A*, located at the center of our Galaxy.

The most significant exact models of black holes are the Schwarzschild solution (1916) and the Kerr solution (1963). These models represent unique stationary vacuum spacetimes that are spherically and axially symmetric, respectively, and asymptotically flat. These foundational metrics have been extensively utilized to explore their intriguing mathematical properties and, importantly, to investigate~(astro)physical phenomena occurring in the vicinity of static and rotating black holes, including their effects on matter and fields, thermodynamics, and quantum phenomena.

Generalizations of these spacetimes to the case of a non-trivial electromagnetic field are known for a long time. These are the Reissner-Nordstr\"{o}m (1916,1918) and Kerr-Newman spacetimes (1965). They are important classes as, according to the uniqueness theorems, these are the only asymptotically flat black holes within Einstein-Maxwell gravity. However, relaxing the conditions of asymptotic flatness may lead to unusual settings with a non-trivial electromagnetic field. Examples of such generalizations were found by Ernst and Wild \cite{ernst1,ernst2} by employing the Harrison transformation. These spacetimes describe black holes, immersed into the Bonnor-Melvin universe \cite{bonnor,melvin}. Modern generalizations can be found in \cite{pinto}. These spacetimes have been for a long time considered as a model of a black hole in a magnetic field (in the Bonnor-Melvin background representing an axially-symmetric magnetic field coupled to gravity). We have to note that under ``background" here and further we understand that if there is no black hole (namely, if the mass parameter $m$ is zero) then the spacetime reduces to Bonnor-Melvin and this notion of ``background" is not related to the asymptotic structure. However, several objections were found against using these spacetimes as a proper physical model of a black hole in a magnetic field. These objections include that (i) these spacetimes are of algebraic type I (meaning that they radiate) \cite{pravda}, (ii) their ergoregions may extend to infinity \cite{Gibbons}, (iii) both charged and uncharged particles cannot escape to infinity in the equatorial plane \cite{Galtsov}.

These facts show that, first of all, if one does not require asymptotical flatness, some solutions, describing a black hole in an external electromagnetic field of the Einstein-Maxwell equations, may exist, and, secondly, that current models of spacetimes in external electromagnetic fields are not sufficient as realistic models. This motivated our work, where we aim to obtain a new class of black hole spacetimes with a non-trivial electromagnetic field, being a counterpart to already known Bonnor-Melvin-based spacetimes. We use several assumptions. Namely, we require that the spacetime remain of algebraic type D (as the Kerr-Newman black hole). Also, we assume that \textit{none} of the principal null directions of the Weyl tensor is aligned with the eigendirections of the Faraday tensor. The reason for this assumption is simple. The case of double alignment corresponds to the already known Pleba\'{n}ski-Demia\'{n}ski solution (and Kerr-Newman black hole as its subcase), while the case when only one direction is aligned does not contain Kerr spacetime as its subcase \cite{Leroy1,Leroy2} (thus is not of our interest).

\section{Twisting black hole with a non-aligned electromagnetic field}

The task of integrating the field equations, even under these simplifying assumptions, is quite complicated. Below we present only the result in the Griffiths-Podolsk\'{y} metric form, presented in Section VII in \cite{ovcharenko1} (full integration of field equations giving this solution can be found in \cite{ovcharenko1})
\begin{align}
    \dd s^2=\dfrac{1}{\Omega^2}\Bigg[&-\dfrac{\mathcal{Q}}{\varrho^2}
    \Big(\dd t-\big[a\sin^2\theta+2(l+\omega\, x_0)(1-\cos\theta)\big]\dd\varphi\Big)^2
    +\dfrac{\varrho^2}{\mathcal{Q}}\,\dd r^2 \nonumber\\
    &+\dfrac{\varrho^2}{\tilde{\mathcal{P}}}\dd\theta^2+\dfrac{\tilde{\mathcal{P}}}{\varrho^2}\sin^2\theta
    \Big(a\,\dd t-\big[(r+r_0)^2+(a+l+\omega\, x_0)^2\big]\dd\varphi\Big)^2\Bigg],\label{GP_metr}
\end{align}
with the electromagnetic 1-form given by
\begin{align}
    \mathbf{A} = \dfrac{1}{4\alpha\,\bar{c}}\,\dfrac{\omega}{a}\Bigg[\,
    &\Omega_{,r}\,\dfrac{ a\,\dd t-[(r+r_0)^2+(a+l+\omega\, x_0)^2]\,\dd\varphi}
        {(r+r_0)+\im\,(a\cos\theta+l+\omega \,x_0)}         \label{A-4}\\
    +&\dfrac{\im\,\Omega_{,\theta}}{\sin\theta}\,\dfrac{\dd t
    -[a\sin^2\theta+2(l+\omega\, x_0)(1-\cos\theta)]\,\dd\varphi}
       {(r+r_0)+\im\,(a\cos\theta+l+\omega \,x_0)}  + \Omega\,\dd\varphi\, \Bigg]+\mathbf{A}_0\,.\nonumber
\end{align}

Here $\mathcal{Q}$ is a quartic function of $r$, $\tilde{\mathcal{P}}$ is quadratic in $\cos\theta$, $\Omega^2$ is quadratic in both $r$ and $\cos\theta$ (foe exact expressions see Sec. VII in \cite{ovcharenko1}), and $\rho^2=(r+r_0)^2+(a\cos\theta+l+\omega x_0)^2$. This solution depends on 7 parameters, namely, the mass parameter $m$, Kerr $a$ and Taub-NUT $l$ twisting parameters, acceleration $\alpha$, complex parameter $c$ and parameter $\beta$, representing duality rotation between two (not-independent) parameters $r_0$ and $x_0$:
\begin{align}
    \omega x_0=A+R\sin\beta,~~~r_0=B+R\cos\beta\,,
\end{align}
where $A,~B,~R$ are specific functions of $m,~a,~l,~\alpha,~|c|$. The parameter $\mathbf{A}_0$ is just gauge fixing constant and can take any value. The twist parameter $\omega$ has similar meaning and is related to rescailing of coordinates and parameters what allows it to take any value (unless $a$ and $l$ are zero).

Despite its complexity, this spacetime is of algebraic type D, and has a non-aligned electromagnetic field. To understand the role of some of the parameters within this solution better, let us consider 2 important subcases.

\subsection{No acceleration $\alpha=0,~|c|=\mathrm{const}$.}
In this case, the metric functions simplify considerably, and one obtains the Kerr-Newman-NUT spacetime. In this case, it appears that the parameters $r_0$ and $x_0$, introduced above, are related to electric and magnetic charges of a black hole. As $r_0$ and $x_0$ are not independent and are related to parameters $|c|$ and $\beta$, we can conclude that in this special case $|c|$ represents the magnitude of the electromagnetic field, created by the charges of a black hole itself, while $\beta$ represents the duality rotation between electric and magnetic charges. 
\subsection{No acceleration $\alpha=0,~2\alpha|c|=B=\mathrm{const}$.}
This limit is different from the one considered above. It can be shown (see Sec. VIII.E in \cite{ovcharenko1}) that in this case the metric becomes, see \cite{podolsky1} (if in addition $r_0=x_0=0$, which can always be chosen by a proper choice of $\beta$)
\begin{align}
    \dd s^2=\dfrac{1}{\Omega^2}\Big[&-\dfrac{{Q}}{\rho^2}
    \big(\dd t-a\sin^2\theta\,\dd\varphi\big)^2
    +\dfrac{\rho^2}{{Q}}\,\dd r^2
    +\dfrac{\rho^2}{{P}}\,\dd\theta^2+\dfrac{{P}}{\rho^2}\sin^2\theta\,
    \big(a\,\dd t-(r^2+a^2)\,\dd\varphi\big)^2\,\Big],\label{Kerr-BR}
\end{align}
where the metric functions are
\begin{align}
{Q} & = \,\big(1+B^2r^2\big)\, \Delta\,,~~~
\Delta = \Big(1-B^2m^2\,\dfrac{I_2}{I_1^2}\Big) r^2-2m\,\dfrac{I_2}{I_1}\,r + a^2
\,,\label{Delta}\\
\rho^2   & = r^2+a^2\cos^2\theta\,,~~~~ \Omega^2 = \,\big(1+B^2r^2\big) - B^2 \Delta \cos^2\theta\,,\label{rho2}\\[0mm]
{P} &= 1 + B^2 \Big(m^2\,\dfrac{I_2}{I_1^2} - a^2 \Big)\cos^2\theta\,,  \label{tilde_P}
\end{align}
with
\begin{align}
    I_1 = 1-\tfrac{1}{2} B^2a^2  \,,\qquad
    I_2 = 1-B^2a^2   \,.\label{I1I2}
\end{align}

The electromagnetic field is given by the 1-form potential
\begin{align}
    \mathbf{A} = \dfrac{{\rm e}^{\im\,\gamma}}{2B}\,&\Big[\,
    \Omega_{,r}\,\dfrac{ a\,\dd t-({r}^{\,2}+a^2)\,\dd\varphi }{r+\im\,a\cos\theta}  +\dfrac{\im\,\Omega_{,\theta}}{\sin\theta}\,\dfrac{ \dd t - a\sin^2\theta\,\dd\varphi}
       {r+\im\,a\cos\theta} + (\Omega-1)\,\dd\varphi\, \Big]\,.\label{A-7}
\end{align}

For $B=0$ one immediately obtains a Kerr black hole, while if $m=0$, then one obtains the Bertotti-Robinson spacetime, representing a coupled system of uniform electromagnetic field and spacetime geometry, as we were able to show in \cite{podolsky1}. This allows us to interpret the parameter $B$ as the strength of the magnetic field, while $\gamma$ is the duality rotation parameter between magnetic and electric field. Also, if $a=0$, then one obtains a subcase of the solution found by Van den Bergh and Carminati \cite{VandenBergh2020}. All these facts allow us to identify our new solution as a Kerr-Bertotti-Robinson black hole (Kerr-BR for short). Several things have to be noted about this solution. First of all, it can be seen that this solution represents a black hole that is \textit{different} from Ernst's black hole. Unlike the Ernst solution, the Kerr-BR black hole is of algebraic type D, and, as we will show, geodesic (and electrogeodesic) particles may easily escape to infinity.

Horizons of this spacetime are defined by zeros of the function $Q$, and are placed at
\begin{align}
r_\pm= \dfrac{m\,I_2 \pm \sqrt{m^2I_2-a^2I_1^2}}{I_1^2-B^2m^2I_2} \,I_1 \,.  \label{rb}
\end{align}
For a zero Kerr parameter $a=0$, this expression strongly simplifies
\begin{align}
    r_h=\dfrac{2m}{1-B^2 m^2}.
\end{align}
For non-zero Kerr parameter $a$, the inner and outer horizons may coincide (giving an extremal black hole) if one chooses the magnetic field to be
\begin{align}
B^2_{\rm extr}  = \frac{2}{a^4}\big(m-\sqrt{m^2-a^2}\,\big)\sqrt{m^2-a^2}\,.  \label{B-extr}
\end{align}

Analysis of the motion of geodesic particles for the non-twisting case shows that the effective potential for equatorial motion is given by
\begin{align}
\hspace{-3mm}
    V(r)=\Big( 1-B^2m^2-\dfrac{2m}{r}\Big)\Big[1+\dfrac{L^2}{m_0^2}\Big(\dfrac{1}{r^2}+B^2\Big)\Big].
    \label{V_expr}
\end{align}
where $L$ is the conserved angular momentum of the barticle, $m_0$ is the mass of a particle.  It can be easily seen that as $r\to \infty$, $V\to (1-B^2m^2)\Big[1+\dfrac{L^2B^2}{m_0^2}\Big]=\rm{const.}$ This is important: the effective potential remains finite (unlike in the case of the Ernst solution). The same property also holds for electrogeodesic particles.

In addition, one can note that the position of the ISCO observer is exactly $r=3r_h$. This relation is the same as for the Schwarzschild spacetime (but note that the horizon radius is changed because of the magnetic field). The case with a non-zero Kerr parameter is more complicated, but the relation of ISCO radius to inner and outer horizons radii is the same as for the Kerr black hole (see \cite{Wang}). 

\section{Conclusions}

Thus, in this work, we have established a new class of spacetimes of algebraic type D with a non-aligned electromagnetic field. It depends on 7 parameters: mass $m$, Kerr $a$, and NUT $l$ twist parameters, acceleration $\alpha$, complex parameter $c$, related to the strength of the non-aligned electromagnetic field, and $\beta$, related to duality rotation between electric and magnetic charges of a black hole itself. 

By considering specific cases, we were able to show that this new class contains the Kerr-Newman-Taub-NUT black holes, and, in addition, a novel Kerr-BR black hole, representing a black hole in a uniform magnetic field. We hope that our solution will be useful in a broad astrophysical investigations.

\section*{Acknowledgments}
This work has been supported by the Czech Science Foundation Grant No. GA\v{C}R 23-05914S and by the Charles University Grant No. GAUK 260325.

\end{document}